%% file: SkSAF_ArXiv_Main.tex
\documentclass[10pt,twoside,twocolumn,english,aps,prl,superscriptaddress,notitlepage,nobibnotes]{revtex4-2}
\usepackage[utf8]{inputenc}
\usepackage[letterpaper]{geometry}
\usepackage{color,xcolor,soul}
\usepackage{verbatim,textcomp}
\usepackage{amstext,amsfonts,amsmath,amssymb}
\usepackage{graphicx}
\usepackage[calcwidth,explicit]{titlesec}
\usepackage{chngcntr}
\usepackage[normalem]{ulem}
\usepackage{nicefrac, xfrac}
\usepackage[unicode=true,pdfusetitle, bookmarks=false, breaklinks=true,pdfborder={0 0 0},pdfborderstyle={},backref=false,colorlinks=true]{hyperref}
\hypersetup{colorlinks=true,citecolor=blue,linkcolor=black,urlcolor=blue}

\makeatletter\renewcommand\frontmatter@abstractwidth{\dimexpr\textwidth-2.5cm\relax}\makeatother

\titleformat{\section}{}{}{0pt}{}
\titleformat{\section}{\bfseries\sffamily\large\filcenter}{\thesection.}{0.2em}{#1}
\titlespacing{\section}{0pt}{0.2ex}{0.2ex}
\titleformat{\paragraph}[runin]{\normalfont\normalsize\bfseries}{}{0pt}{}
\titlespacing*{\paragraph}{0em}{0ex}{0.3em}[]

\setcitestyle{super}

\renewcommand\thesection{\Alph{section}}
\counterwithout{paragraph}{subsubsection}

\AtBeginDocument{
\renewcommand{\ref}[1]{\autoref{#1}}

}

\begin{document}
\title{Zero Field Antiferromagnetically Coupled Skyrmions and their \\Field-Driven Uncoupling in Composite Chiral Multilayers\smallskip{}
}
\author{May Inn Sim}\thanks{These authors contributed equally to this work}
\affiliation{Department of Physics, National University of Singapore (NUS), 117551 Singapore}
\author{Dickson Thian}\thanks{These authors contributed equally to this work}
\affiliation{Institute of Materials Research \& Engineering, Agency for Science, Technology \& Research (A{*}STAR), 138634 Singapore}
\author{Ramu Maddu}
\affiliation{Institute of Materials Research \& Engineering, Agency for Science, Technology \& Research (A{*}STAR), 138634 Singapore}
\author{Xiaoye Chen}
\affiliation{Institute of Materials Research \& Engineering, Agency for Science, Technology \& Research (A{*}STAR), 138634 Singapore}
\author{Hang Khume Tan}
\affiliation{Institute of Materials Research \& Engineering, Agency for Science, Technology \& Research (A{*}STAR), 138634 Singapore}
\author{Chao Li}
\affiliation{Institute of Materials Research \& Engineering, Agency for Science, Technology \& Research (A{*}STAR), 138634 Singapore}
\author{Pin Ho}
\affiliation{Institute of Materials Research \& Engineering, Agency for Science, Technology \& Research (A{*}STAR), 138634 Singapore}
\author{Anjan Soumyanarayanan}\email{anjan@imre.a-star.edu.sg}
\affiliation{Institute of Materials Research \& Engineering, Agency for Science, Technology \& Research (A{*}STAR), 138634 Singapore}
\affiliation{Department of Physics, National University of Singapore (NUS), 117551 Singapore}

\begin{abstract}
\noindent Antiferromagnetic (AF) skyrmions are topological spin structures with fully compensated, net-zero magnetization.
Compared to their ferromagnetic (FM) skyrmion counterparts, their reduced stray field and enhanced electrical response can enable linear, high-throughput current-driven motion.
However, their bubble-like character in conventional bilayer AFs limits their stability to fluctuations, leading to deformation and annihilation.
Here we present the engineering of a composite AF chiral multilayer, wherein the interplay of AF and FM interlayer couplings generates compensated skyrmions with compact structures. 
High-resolution magnetic imaging and micromagnetic simulations show that the internal exchange field stabilizes AF skyrmions at zero external field with characteristics comparable to FM counterparts at 100 mT.
Quantitative analyses establish their decoupling above the exchange field, yielding independent, spatially segregated textures in constituent chiral layers.
This work provides a foundation to develop AF spin-textures with enhanced immunity, compatible with efficient readout and manipulation, with relevance to unconventional computing.
\end{abstract}
\maketitle

\noindent \input{SkSAF_ArXiv_IntroStack} 
\noindent \input{SkSAF_ArXiv_ResultsConc} 
\noindent \begin{center}
{\small{}\rule[0.5ex]{0.4\columnwidth}{0.5pt}}{\small\par}
\par\end{center}
\noindent \input{SkSAF_ArXiv_Methods}

\noindent \begin{center}
{\small{}\rule[0.5ex]{0.4\columnwidth}{0.5pt}}{\small\par}
\par\end{center}

\noindent \textsf{\textbf{\small{}Acknowledgments.}}\
\noindent {\small{}We acknowledge inputs from Kevin Masgrau, Michael Tran, and S. Goolaup, as well as the support of the National Supercomputing Centre (NSCC), Singapore for computational resources. This work was supported by the SpOT-LITE programme (Grant Nos. A1818g0042, A18A6b0057), funded by Singapore's RIE2020 initiatives.}

\phantomsection\addcontentsline{toc}{section}{\refname}
\bibliography{SkSAF}

\noindent \begin{center}
{\small{}\rule[0.5ex]{0.6\columnwidth}{0.5pt}}{\small\par}
\par\end{center}
\end{document}

%% file: SkSAF_ArXiv_IntroStack.tex
\section{Introduction} \label{Introduction}

\paragraph{Motivation}
Magnetic skyrmions are topologically wound, nanoscale spin structures that form within a uniform magnetic background \cite{Nagaosa2013, Fert2017}.
First discovered in chiral magnets \cite{Muhlbauer2009, Yu2010}, 
they can be generated at room temperature in multilayer thin films \cite{Moreau-Luchaire2016, Boulle2016, Soumyanarayanan2017}.
Their current-driven dynamics in racetracks \cite{Woo2016, Jiang2015, Jiang2017, Litzius2017, Song2024} as well as efficient readout and switching in tunnel junctions \cite{Chen2024, Larranaga2024, Zhao2024} are promising for emerging computing applications \cite{Fert2017, Grollier2020, Back2020}. 
However, several prevailing attributes of conventional ferromagnetic (FM) skyrmions inhibit the realization of practical devices. 
Their equilibrium size and resilience to perturbations are constrained, especially at zero magnetic field (ZF), by their intrinsic stray field \cite{Buttner2018, Yagil2018, Dovzhenko2018} and instability to elongation \cite{Leonov2016, Tan2020}. 
Meanwhile, their electrical mobility is limited by their topological charge, which introduces both transverse deflection \cite{Jiang2017, Litzius2017, Zeissler2020, Tan2021} and additional damping contributions \cite{Buttner2018}.
These limitations can be circumvented by compact antiferromagnetic (AF) skyrmions,  whose minimal stray field and zero effective topological charge can enable enhanced resilience and mobility \cite{Barker2016, Zhang2016b}.   

\paragraph{Literature Review}
Promising skyrmion characteristics observed in ferrimagnets \cite{Woo2018, Hirata2019, Mandru2020, Quessab2022} have motivated efforts to develop fully-compensated AF textures. 
Intrinsic chiral antiferromagnets host a zoo of myriad textures\cite{Jani2021}, and exchange-bias imprinting of ferromagnets onto conventional antiferromagnets yields compensated skyrmions with hybrid chirality \cite{Rana2021, He2024} -- whose functionalities remain to be established. 
Meanwhile, practically most promising are synthetic antiferromagnets (SAFs), which can host oppositely polarized skyrmions in adjacent chiral layers \cite{Duine2018, Wang2023}. 
SAF skyrmions may retain known benefits of FM skyrmions, with enhancements from AF compensation \cite{Zhang2016b}. 
Several works on bilayer SAFs show compensated skyrmion bubbles \cite{Legrand2020, Chen2020, Finco2021, Juge2022, Barker2024}, nucleation via electrical and optical techniques\cite{Juge2022, Mallick2024}, and near-linear current-driven dynamics \cite{Dohi2019, Pham2024}. 
Notably, their compensated topological charge has been recently employed to realize uninhibited skyrmion mobility \cite{Pham2024}. 
However, key challenges concerning skyrmion stability to deformation, annihilation, and pinning remain to be addressed \cite{Pham2024}.
These limitations are inextricably linked to the inherent "bubble" character of skyrmions within the individual chiral layers of bilayer SAFs.

\paragraph{Results Summary}
We address this bottleneck by designing an Ir/Fe/Co/Pt-based composite chiral SAF multilayer, which hosts chiral AF textures by synergistically combining FM and AF interlayer exchange couplings (IECs).
Quantitative comparison of spin-textures in SAF stacks with FM counterparts via magnetic force microscopy (MFM), magnetometry, and micromagnetic simulations, cements their stray-field compensated compact structure. 
While FM-IEC enables inherent textural stability within each constituent chiral stack, AF-IEC between chiral stacks provides a large exchange field ($\sim$100 mT), facilitating the presence of SAF skyrmions at ZF.  
With increasing field, they decouple and yield independent, spatially segregated textures in the constituent chiral multilayers. 
These results provide a platform for developing robust SAF skyrmions compatible with all-electrical control \cite{Chen2024}.

\section{Stack Design and Properties\label{sec:StackDesignAndProperties}}

\begin{figure}[htb]
\noindent \begin{centering}
\includegraphics[width=3.3in]{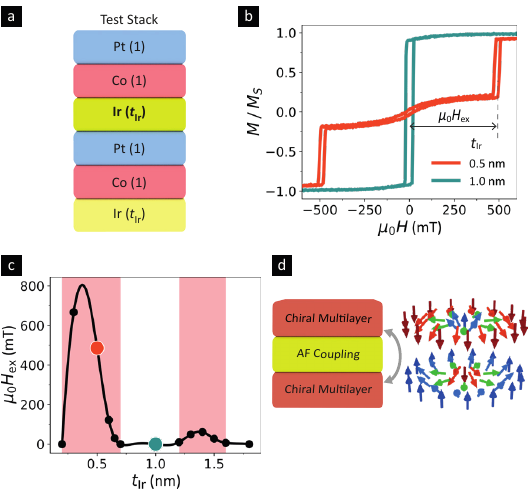}
\par\end{centering}
\noindent \caption[Modulating IEC between Chiral Stacks.]{\textbf{IEC between Chiral Stacks.} 
\textbf{(a)} Schematic of bilayer Ir/Co/Pt stack (layer thickness in nm in parenthesis) used to study interlayer exchange coupling (IEC) with Ir thickness, $t_{\rm Ir}$. 
\textbf{(b)} Normalised out-of-plane (OP) magnetization hysteresis $M(H)$ loops for representative $t_{\rm Ir}$ = 0.5 nm (orange, antiferromagnetic (AF)-IEC) and $t_{\rm Ir}$ = 1.0 nm (green, ferromagnetic (FM)-IEC). Dotted line indicates AF-IEC exchange field $\mu_0 H_{\rm ex}$ (defined in text). 
\textbf{(c)} Evolution of $\mu_0H_{\rm ex}$ with $t_{\rm Ir}$ over 0.2 – 1.8 nm. Shaded pink regions correspond to AF-IEC ($H_{\rm ex}>0$). 
\textbf{(d)} Left: schematic of synthetic AF coupled multilayers, each hosting skyrmions. Right: schematic of resulting SAF skyrmion (arrows indicate magnetic moments).
\label{fig:SAF-StackDesign}}
\end{figure}

\paragraph{Oscillatory IEC \& Design Principle}
In multilayer films, two magnetic layers separated by a non-magnetic spacer exhibit IEC due to the interlayer Ruderman-Kittel-Kasuya-Yosida (RKKY) interaction \cite{Bruno1995}. 
RKKY-induced IEC typically presents damped oscillations with spacer thickness.
Moreover, it can be ambipolar for some metals, e.g., Ru, Ir etc., enabling varying strengths of FM (parallel) and AF (anti-parallel) alignment \cite{Parkin1991}. 
Here, Ir is chosen as the IEC-mediating spacer \cite{Yanagihara1999} as it also generates sizable interfacial Dzyaloshinskii-Moriya interaction, required for chiral spin textures \cite{Moreau-Luchaire2016, Soumyanarayanan2017, Chen2022}. 

\paragraph{IEC Variation with Ir Thickness}
To characterize IEC evolution, multilayer stacks comprising two repetitions of Ir($t_{\rm Ir}$)/Co(1)/Pt(1) (thickness in nm in parenthesis, \ref{fig:SAF-StackDesign}a) were deposited for varying Ir thickness, $t_{\rm Ir}$ ($0.2-1.8$ nm).
Magnetometry measurements consistently evidence perpendicular anisotropy, and out-of-plane (OP) magnetization hysteresis loops ($M(H)$) exhibit two distinct shapes across varying $t_{\rm Ir}$ (\ref{fig:SAF-StackDesign}b).
For $t_{\rm Ir}\sim 1$ nm, they have typical FM (near-rectangular) character, with high remanence (\ref{fig:SAF-StackDesign}b: green). 
For $t_{\rm Ir} < 0.8$ nm (\ref{fig:SAF-StackDesign}b: orange), $M$ is near-zero for $|H|< H_ {\rm ex}$, due to anti-parallel alignment of the Co layers, and near-$M_{\rm s}$ for larger fields. 
Here $M_{\rm s}$ is the saturation magnetization (i.e., at $H_{\rm s}$), and the exchange field, $H_ {\rm ex}$, is defined as $M(H_{\rm ex})=M_ {\rm s}/2$.   
\ref{fig:SAF-StackDesign}c shows the variation of $\mu_0H_ {\rm ex}$ for $t_ {\rm Ir}$ over $0.2-1.8$ nm, with characteristic oscillatory behaviour \cite{Parkin1991}. 
AF-IEC ($H_ {\rm ex}>0$) is observed for $t_{\rm Ir}$ over $0.2-0.8$ nm, and also over $1.2-1.5$ nm (\ref{fig:SAF-StackDesign}c: pink regions), albeit with reduced magnitude for the latter due to the spatial decay of IEC. 
These trends are consistent with prior reports on Co/Ir-based multilayers \cite{Yanagihara1999, Yakushiji2017}. 

\paragraph{Composite Stack Design}
Combining such AF coupling with FM-IEC can extend compensated stacks from bilayers to multilayers \cite{Bran2009, Duine2018}.
Here, we posit that chiral spin textures inherently generated in near-identical constituent chiral stacks \cite{Chen2023} can be AF-coupled to generate compact SAF textures (\ref{fig:SAF-StackDesign}d, SI $\S$S1).
Our base stack – Ir/Fe/Co/Pt – is an established  host of nanoscale N\'eel skyrmions with tunable properties \cite{Soumyanarayanan2017, Chen2022}.
The chiral multilayer comprises three repetitions of Ir(1)/Fe(0.2)/Co(1.2)/Pt(0.8) (see Methods) to ensure robust skyrmion formation \cite{Chen2023}, while minimizing complexities arising from interlayer dipolar coupling effects \cite{Hellwig2007}.
The chosen FM thicknesses optimize imaging contrast, while those of Pt(0.8) and Ir(1.0) ensure a strongly coupled multilayer. 
Next, two such multilayers are coupled by an Ir spacer  with two distinct thicknesses (\ref{fig:SAF-MH_MFM}a) –  generating AF-IEC ($t_{\rm Ir} =$ 0.5 nm, identified as AFS), and FM-IEC ($t_{\rm Ir} =$ 1.0 nm: FMS), respectively. 
The composite stack sandwiches the critical Ir spacer with identical Co(1.0) layers that ensure coherent coupling.

%% file: SkSAF_ArXiv_ResultsConc.tex
\section{Texture Evolution in Composite Stacks}\label{sec:FieldEvolution}

\begin{figure*}[htb]
\noindent \begin{centering}
\includegraphics[width=6.9in]{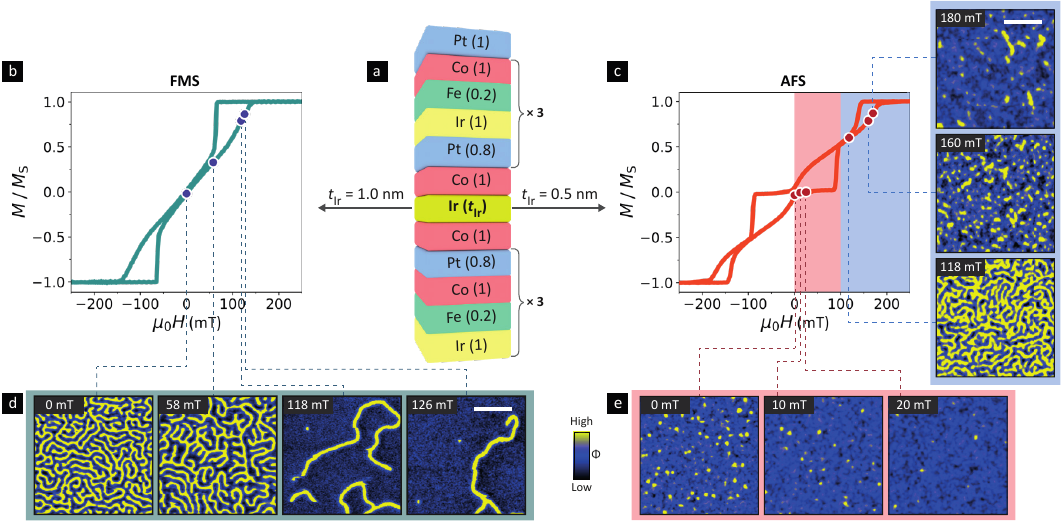}
\par\end{centering}
\noindent \caption[Imaged Field Evolution of FMS and AFS Stacks.]{\textbf{Imaged Field Evolution of FMS and AFS Stacks.} 
\textbf{(a)} Schematic structure of the two Ir/Fe/Co/Pt-based composite multilayer stacks, identified as FMS ($t_\mathrm{Ir} = 1.0$ nm, left) and AFS ($t_\mathrm{Ir} = 0.5$ nm, right) respectively. 
\textbf{(b-c)} Normalized OP $M(H)$ loops of (b) FMS ($t_\mathrm{Ir} = 1.0$ nm) and (c) AFS ($t_\mathrm{Ir} = 0.5$ nm). Shaded circles indicate MFM field points. \textbf{(d-e)} Representative MFM images of spin textures of (d) FMS and (e) AFS (scalebar: 1 \textmu m) at indicated OP fields.}
\label{fig:SAF-MH_MFM}
\end{figure*}

\paragraph{MFM of FM Stack}
\ref{fig:SAF-MH_MFM}b-e show results from magnetometry and field-dependent MFM imaging of the composite stacks. 
For the FMS (\ref{fig:SAF-MH_MFM}b), the $M(H)$ loop is sheared, and MFM (\ref{fig:SAF-MH_MFM}d) expectedly shows a labyrinthine ZF state. 
With increasing $\mu_0H$, it evolves into discrete stripes and sparse skyrmions enroute to saturation -- a transition well-observed for chiral multilayers \cite{Woo2016, Moreau-Luchaire2016, Soumyanarayanan2017}.
Contrastingly, the $M(H)$ loop of the AFS (\ref{fig:SAF-MH_MFM}c) exhibits qualitatively distinct evolution, with a ZF plateau and two distinct sheared regions (shaded pink, blue), separated by a jump at $H_{\rm ex}$, enroute to $H_{\rm s}$. 
The reduced ZF plateau for AFS ($\mu_0 H_{\rm ex} \sim110$ mT) c.f. Ir/Co/Pt bilayer (\ref{fig:SAF-StackDesign}b: $\mu_0 H_{\rm ex}\sim400$ mT), is attributed to the composite stack, and its lower $K_{\rm eff}$ \cite{Yanagihara1999, Yakushiji2017}. 
Meanwhile, MFM also reveals two distinct field-evolution regimes, with the near-ZF region (\ref{fig:SAF-MH_MFM}e) showing sparse, mostly skyrmionic textures, which disappear into the uniform background at $\sim$30 mT. 
Above $\sim$110 mT ($\sim$$\mu_0 H_{\rm ex}$, \ref{fig:SAF-MH_MFM}e: right), a labyrinthine state suddenly reappears, and evolves into stripes, then skyrmions, and finally to a uniform state at $\sim$180 mT ($\sim$$\mu_0 H_s$). 

\begin{figure*}[htb]
\noindent \begin{centering}
\includegraphics[width=6.9in]{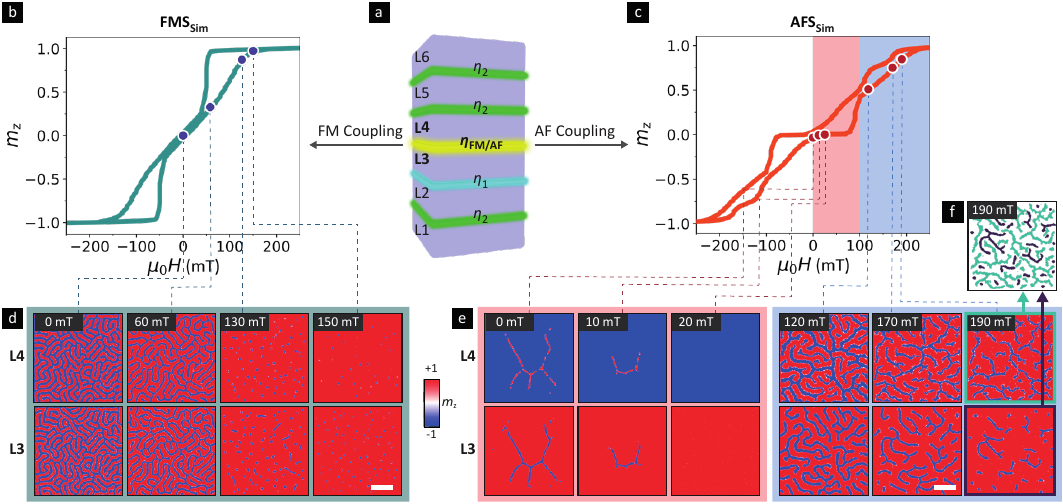}
\par\end{centering}
\noindent \caption[Simulated Field Evolution of FMS and AFS Stacks.]{\textbf{Simulated Field Evolution of FMS and AFS Stacks.} 
\textbf{(a)} Schematic of simulated composite stack with six effective layers (L1-6). IECs within (between) the two chiral stacks (L1-3, L4-6) are denoted as $\eta_{\rm 1,2}$ ($\eta_{\rm FM,AF}$) respectively. 
\textbf{(b-c)} Simulated normalized OP $m_z(H)$ loops of (b) $\mathrm{FMS_{Sim}}$ ($\eta_{\rm FM}$) and (c) $\mathrm{AFS_{Sim}}$ ($\eta_{\rm AF}$). 
\textbf{(d-e)} Representative magnetization images of (d) FMS and (e) AFS for layers L3, L4 (scalebar: 0.5 \textmu m) at indicated OP fields (shaded circles), depicting spin-texture evolution.
\textbf{(f)} Superimposed image of textures in layers L3 (purple) and L4 (green) for $\mathrm{AFS_{Sim}}$ at $\mu_{\rm 0}H$ = 190 mT, showing their distinct positions.
\label{fig:SAF-MH_Sim}}
\end{figure*}

\paragraph{Simulations Setup \& Methodology}
To lay out a phenomenological picture, we performed grain-free micromagnetic simulations of a minimum-model, six-layer composite stack designed to qualitatively emulate the key features of experimentally studied AFS and FMS structures (\ref{fig:SAF-MH_Sim}a). The simulations employed established recipes and parameters from related prior works (see Methods, SI §S2) \cite{Chen2022, Bottcher2023, Chen2023}.
The IECs between successive FMs were modeled using three distinct parameters ($\eta_{\rm 1}, \eta_{\rm 2}, \eta_{\rm FM/AF}$), the distinct heavy metal (HM) spacers \cite{Parkin1991, Yakushiji2017, Legrand2020, Chen2023}. 
Notably, opposite signs of $\eta_{\rm FM,AF}$ generate the distinct behaviours of $\mathrm{AFS_{Sim}}$ ($\eta_{\rm AF}>0$) and $\mathrm{FMS_{Sim}}$ ($\eta_{\rm AF}<0$) respectively \cite{Yanagihara1999, Yakushiji2017}.

\begin{figure*}[htb]
\noindent \begin{centering}
\includegraphics[width=6.9in]{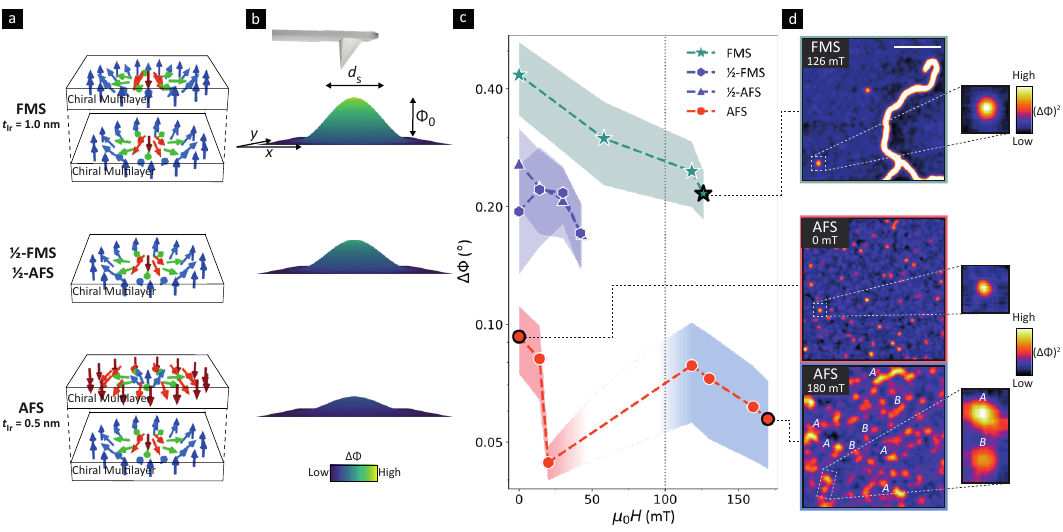}
\par\end{centering}
\noindent \caption[Magnetic Phase Contrast (MPC) Evolution of Spin Textures.]{\textbf{Magnetic Phase Contrast (MPC) Evolution of Spin Textures.}
\textbf{(a-b)} Schematics of expected (a) orientation of skyrmionic spin textures (arrows) and (b) their MPC ($\mathrm{\Delta\Phi}$) profile across FMS (top), $\nicefrac{1}{2}$-FMS, -AFS (middle), and AFS (bottom) samples (top inset: schematic of MFM tip), and skyrmion fit parameters (amplitude, $\mathrm{\Phi_0}$, diameter, $d_{\rm S}$). 
\textbf{(c)} Field evolution of average MPC amplitude, $\langle\mathrm{\Delta\Phi_0}\rangle (H)$, for the four samples. Each point averages all identifiable spin-textures over imaged fields-of-view (see Methods, SI $\S$S3), shaded regions denote standard deviation. 
\textbf{(d)} Representative MFM images (highlighted field-points in (c), scalebar: 1 \textmu m, colour-bar: $\mathrm{\Delta\Phi}^2$) of isolated spin textures for FMS at near-$H_{\rm s}$ (top), AFS at ZF (middle), and AFS at near-$H_{\rm s}$ (bottom). Right insets: zoom-in MFM images of typical skyrmions for each case (bottom: two characteristic intensities, A, B).
\label{fig:SAF-PhaseContrast}}
\end{figure*}

\paragraph{Simulation - Experiment Comparison}
The OP hysteresis loops, $m_{\rm z}(H)$, of the two simulated stacks (\ref{fig:SAF-MH_Sim}b-c) evidence distinct evolutions, qualitatively similar to the respective experimental results (\ref{fig:SAF-MH_MFM}b-c).
While the $\mathrm{FMS_{Sim}}$ loop (\ref{fig:SAF-MH_Sim}b) is typically sheared, the $\mathrm{AFS_{Sim}}$ loop (\ref{fig:SAF-MH_Sim}c) resembles the concatenation of two FMS loops at $\mu_0H_{\rm ex} \simeq 100$ mT.
Upon ascertaining the validity of the simulated stacks, we turn to the microscopic magnetization evolution, wherein simulations provide layer-wise information on spin texture character (\ref{fig:SAF-MH_Sim}d-e: layers L3, L4 adjacent to $\eta_{\rm FM/AF}$).
While $\mathrm{FMS_{Sim}}$ exhibits prototypical stripes and skyrmions with identical magnetization across layers (\ref{fig:SAF-MH_Sim}d), $\mathrm{AFS_{Sim}}$ exhibits two distinct texture evolution trends demarcated by $\mu_0H_{\rm ex}$. 
At and around ZF (\ref{fig:SAF-MH_Sim}e), $\mathrm{AFS_{Sim}}$ shows isolated stripe textures with compensated (SAF-like) character, i.e., L3 and L4 magnetizations are coincident and anti-aligned. 
These compensated textures shrink and disappear at $\sim$30 mT ($\ll\mu_0H_{\rm ex}$).
At $\mu_0H_{\rm ex}\simeq120$ mT ($>$$\mu_0H_{\rm ex}$), spin textures abruptly reappear as a labyrinthine configuration, however with L3 and L4 magnetizations aligned and with distinct spatial configurations.
With increasing field, the two layers exhibit distinct stripe and skyrmion configurations, which disappear at $\sim$200 mT ($\sim$$\mu_{\rm 0}H_{\rm s}$). 
Meanwhile, the observed variation in simulated layer-dependent texture intensities, arising from known limitations of the grain-free minimal model (SI §S2), motivate future efforts to achieve quantitative consistency with experiments.

\paragraph{Simulation Insights}
The simulations elucidate several aspects of spin texture evolution in the composite AFS. 
First, the MFM-imaged round textures at ZF are SAF-coupled skyrmions, and are visible due to their imperfectly compensated inhomogeneous stray fields. 
Second, there is clear dichotomy in texture character around $H_{\rm ex}$. 
At higher fields ($H > H_{\rm ex}$), the two SAF-coupled stacks should each host distinct textures with aligned magnetization. 
In fact, as seen in \ref{fig:SAF-MH_Sim}f, these texture configurations do not exhibit any overlap, and are in fact spatially segregated and seemingly anti-correlated. 
The observed AFS phenomenology, arising from the interplay of IEC, bias field, and texture character, motivates quantitative imaging studies.

\section{Phase Contrast \& Texture Deconvolution\label{sec:MFM-analysis}}

\paragraph{MPC Principle}
The magnetic phase contrast (MPC), $\mathrm{\Delta\Phi}$, detected by the MFM tip (magnetization, $\vec{M}_{\rm tip}$) is given as \cite{Kiselev2008, Kazakov2019}
\noindent
\begin{equation}
\Delta\Phi \sim \int_{V_{\rm tip}}\Vec{M}_{\rm tip}(\vec{r})\cdot\frac{\partial^2\Vec{H}_S(\Vec{r})}{\partial z^2} 
 ,\label{eq:PC_Form}
\end{equation}
where $\vec{H}_{\rm S}$ is the stray field from the sample, and the integral is over tip volume, $ V_{\rm tip}$. 
Previous works have employed MPC to study the helicity, spatial extent, and magnetic parameters of spin textures \cite{Yagil2018, Bacani2019, Legrand2020, Mandru2020}.
For a multilayered film, $\vec{H}_{\rm S}$ is a vector superposition of stray fields from each magnetic layer, which would sum constructively (destructively) for FM- (AF-) coupled stacks.
Meanwhile, the larger stray field from near-surface layers can enable MPC quantification of magnetization-compensated SAF skyrmions \cite{Legrand2020}, in contrast to their null signal in transmission techniques \cite{Cheong2020}. 
Therefore, MPC can provide a detailed picture of the evolution of spin texture character. 

\paragraph{MPC Analysis Method}
Quantitative MPC analysis requires a principled approach to disentangle tip, sample, and field-dependent contributions (\ref{eq:PC_Form}). 
Accordingly, copies of AFS and FMS samples were carefully etched to remove the top half of the stack above the Ir-coupling layer (see \ref{fig:SAF-MH_MFM}a). 
The resulting $\nicefrac{1}{2}$-AFS and $\nicefrac{1}{2}$-FMS samples with nominally identical magnetic properties are suited for validation (see Methods, SI $\S$S1,2). 
Across these samples (\ref{fig:SAF-PhaseContrast}a-b: schematics), we expect MPC to be highest for FMS (top), followed by $\nicefrac{1}{2}$-FMS and $\nicefrac{1}{2}$-AFS (middle), and lowest for AFS (bottom).
To ensure objective comparison, the entire dataset (\ref{fig:SAF-PhaseContrast}c) was acquired using a single MFM tip, with identical scan conditions across samples and magnetic fields. 
Subsequently, all identified textures across MFM images were iteratively fit to determine their MPC amplitude, $\mathrm{\Phi_0}$, and texture width {(see Methods, SI $\S$S3)}. 

\paragraph{MPC Analysis Trends}
\ref{fig:SAF-PhaseContrast}c shows the variation of ensemble-averaged MPC, $\langle\mathrm{\Phi_0}\rangle (H)$, with OP field ($\mu_0H$) quantified over large, statistically significant fields-of-view for the four samples.
Excepting AFS, the samples exhibit similar field-trends with distinct amplitudes, separated by 5-10 times the magnitude of individual statistical spreads.
Expectedly, the MPC is highest for FMS (\ref{fig:SAF-PhaseContrast}c: top, $\langle\mathrm{\Phi_0}\rangle^{\rm ZF}\approx0.4^\circ$), followed by near-identical values for $\nicefrac{1}{2}$-FMS and $\nicefrac{1}{2}$-AFS, (\ref{fig:SAF-PhaseContrast}c: middle, $\langle\mathrm{\Phi_0}\rangle^{\rm ZF}\approx0.2^\circ$), validating the experimental methodology. 
Next, $\langle\mathrm{\Phi_0}\rangle (H)$ reduces monotonically ($30-50\%$) likely due to field-induced texture shrinking and tip magnetization alignment effects -- also evidenced in micromagnetic simulations (SI $\S$S2), and in prior works on microscale spin-textures \cite{Bran2009}. 
Finally, for AFS, the MPC is substantially lower (\ref{fig:SAF-PhaseContrast}c: bottom, $\langle\mathrm{\Phi_0}\rangle^{\rm ZF}<0.1^\circ$), confirming their stray-field-compensated SAF character at ZF. 
Crucially, while $\langle\mathrm{\Phi_0}\rangle (H)$ initially decreases, it sharply jumps upward at $\sim$100 mT ($\sim$$\mu_{\rm 0}H_{\rm ex}$), before decreasing to saturation.

\paragraph{MPC Analysis Insights}
The abrupt jump in the MPC evolution of AFS evidences a transition in texture character, consistent with their above $H_{\rm ex}$ uncoupling between chiral stacks expected from simulations (\ref{fig:SAF-MH_Sim}f). 
The uncompensated stray field of uncoupled textures should yield MPC larger than for SAF textures (AFS: near-ZF), yet lower than for FMS textures. 
Indeed, $\langle\mathrm{\Phi_0}\rangle (H)$ of AFS for $H>H_{\rm ex}$ smoothly extrapolates to that of $\nicefrac{1}{2}$-AFS/FMS stacks, indicating their qualitative similarity. 
Meanwhile, closer inspection of MFM images (\ref{fig:SAF-PhaseContrast}d: bottom) reveals two distinct sets of MPCs within the same field-of view. 
Statistical analysis of $H>H_{\rm ex}$ spin-texture ensembles consistently shows a bimodal (two-peaked) MPC distribution (SI $\S$S3), in contrast to Gaussian (single-peak) distributions for conventional spin-textures. 
Simulations predict distinct, spatially segregated spin textures (\ref{fig:SAF-MH_Sim}e-f) in the two component stacks, which should manifest in MFM with different MPCs. 
Thus, we conclude that higher MPC textures (\ref{fig:SAF-PhaseContrast}d: bottom, “A”) form in the top half-stack, while lower MPC textures (\ref{fig:SAF-PhaseContrast}d: bottom, “B”) are in the bottom half-stack. 
Future minor-loop magnetometry-imaging studies of the AFS may shed valuable insights on the formation mechanisms and transitions of SAF-coupled and uncoupled textures \cite{Hellwig2007, Tan2020, Chen2022a}. 

\begin{figure}
\noindent \begin{centering}
\includegraphics[width=3.3in]{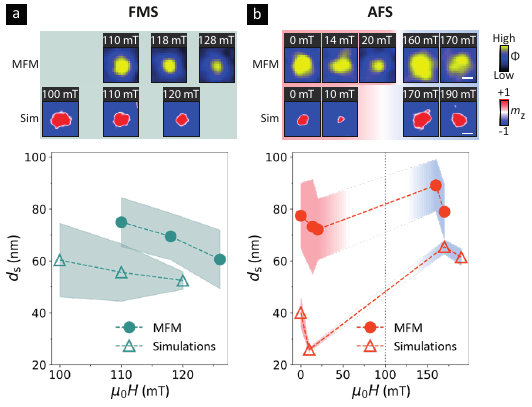}
\par\end{centering}
\noindent \caption[Field Evolution of Skyrmion Sizes.]{\textbf{Field Evolution of Skyrmion Sizes.}
Measured size of isolated skyrmions, $\mu_{\rm 0}H$, across MFM experiments (round markers) and simulations (triangle markers) for \textbf{(a)} FMS and \textbf{(b)}AFS samples. Each data point averages all identifiable skyrmions over imaged fields-of-view, shaded regions denote standard deviation. 
Top insets (scalebar: 50 nm) show zoom-in images of representative skyrmions for MFM (top) and simulations (bottom).
\label{Fig:SAF-Size}}
\end{figure}

\paragraph{Skyrmion Size Trends}
Finally, we compare in \ref{Fig:SAF-Size} the evolution of average skyrmion size, $d_{\rm S}(H)$ for AFS and FMS samples (details in SI $\S$S3).
The two stacks exhibit distinct trends, with qualitative consistency between experiments and simulations. 
For FMS (\ref{Fig:SAF-Size}a), skyrmions form near saturation ($\sim$100 mT), and $d_{\rm S}$ ($\sim$80 nm) decreases by $\sim$20$\%$ over a narrow field range, in line with literature \cite{Moreau-Luchaire2016, Soumyanarayanan2017}. 
Meanwhile, for AFS, $d_{\rm S}$ first decreases over the low-field SAF-coupled regime, then abruptly jumps up for the uncoupled regime ($H>H_{\rm ex}$), and decreases again to saturation. 
These AFS and FMS field trends mirror those for the MPC amplitudes (\ref{fig:SAF-PhaseContrast}c), evidencing their shared origin \cite{Bran2009}. 
Notably, while AFS and FMS employ identical chiral stacks, the near-ZF SAF skyrmions are directly comparable to FM skyrmions at the highest fields ($\mu_0H \sim100$ mT).
This underscores the profound implications of AF-IEC on skyrmion stability, energetics, and kinetics \cite{Chen2023}. 
Finally, the reported SAF skyrmion sizes compare well with literature \cite{Legrand2020, Finco2021, Juge2022, Pham2024, Barker2024}.
Future works can explore composite SAF stacks with strongly chiral layers (e.g., \cite{Soumyanarayanan2017, Chen2022}) to develop SAF skyrmions with further reduced size and enhanced stability.

\section{Conclusion \label{sec:Summary}}

\paragraph{Summary}
In summary, we have established a composite multilayer platform wherein the character and evolution of chiral spin textures is demonstrably modulated by the interplay of stack-tunable AF- and FM-IECs, and by magnetic field.
The AFS stack generates robust SAF-coupled skyrmions at ZF, whose internal stray field compensation results in strongly diminished contrast compared to FM textures.
Strikingly, the internal effective field from AF-IEC ensures that the SAF skyrmions at ZF are compact, with sizes comparable to FM counterparts at over 100 mT. 
Moreover, the composite stack decouples above the AF-IEC field ($H_{\rm ex}$), yielding individual, spatially segregated spin-textures in each half-stack. 

\paragraph{Impact}
The presented realization of SAF-coupled ZF skyrmions with reduced size and enhanced stability has immediate relevance to devices employing skyrmions and beyond. 
First, while skyrmions in bilayer SAFs exhibit fast, linear motion \cite{Pham2024}, there are inevitable fluctuations in properties due to their bubble-like nature. 
In contrast, compact SAF skyrmions -- strengthened additionally by FM-IEC \cite{Chen2023} --  should be endowed with greater resilience to perturbations \cite{Tan2021}. 
This should enable more coherent and robust dynamics with racetrack devices \cite{Parkin2008, Fert2017}.
Second, uniform SAF stacks are being increasingly explored as the active layer for magnetic tunnel junctions (MTJs) with enhanced immunity \cite{Cuchet2016}. 
Incorporating a SAF skyrmion stack within two-terminal MTJs can enable multiple, robust electrical states, with facile manipulation \cite{Chen2024}. 
Lastly, the hitherto unexplored skyrmion coupling crossover ($H\sim H_{\rm ex}$), governed by AF- and FM-IECs \cite{Legrand2020, Chen2023}, presents a rich, tunable spin-texture phase space \cite{Mandru2020,Jani2021,Grelier2022, Bhukta2024}.
The ability to propagate quasi-independent skyrmions in each layer, or to control propagation in one layer via the other, offers lucrative prospects for stacked racetracks. 
Such devices may be employed for reconfigurable logic and unconventional computing \cite{Fert2017}, in particular within probabilistic and reservoir architectures \cite{Finocchio2021}.

%% file: SkSAF_ArXiv_Methods.tex
\section{Methods}

\textbf{Film Deposition.} Multilayer thin films were deposited on thermally oxidized Si/SiO$_2$ wafers by DC magnetron sputtering at room temperature with base pressure of 2×10$^{-8}$ Torr, using a Chiron™ PVD system manufactured by Bestec GmbH \cite{Soumyanarayanan2017, Tan2020}. 
The primary stacks studied have compositions: Ta(3)/Pt(5) / [Ir(1)/Fe(0.2)/Co(1)/Pt(0.8)]$_3$ / Co(1)/Ir($t_{\rm Ir}$)/Co(1) / [Pt(0.8)/Ir(1)/Fe(0.2)/Co(1)]$_3$ / Pt(3), with $t_{\rm Ir}$= 0.5 (AFS) and 1.0 (FMS) respectively. 
Bilayer [Ir($t_{\rm Ir}$)/Co(1)/Pt(1)]$_2$ stacks, used to determine AF-IEC (\ref{fig:SAF-StackDesign}), were deposited using identical buffer and capping layers. 
Finally, $\nicefrac{1}{2}$-FMS and $\nicefrac{1}{2}$-AFS stacks were fabricated by etching the primary stacks with precise end-point detection (see SI $\S$S1).

\textbf{Magnetometry.} $M(H)$ loops were measured on diced samples using a MicroMag™ alternating gradient magnetometer. 
The measured magnetic properties of all stacks are detailed in SI $\S$S1, and consistent with our previous works \cite{Tan2020, Chen2022}.

\textbf{MFM Experiments} were performed using a Bruker Dimension Icon™ 3100 atomic force microscope, using Co-alloy coated SSS-MFMR tips from NanoSensors™ (magnetization $\sim$80 emu/cm$^3$, diameter $\sim$30 nm). 
Before each imaging dataset, the tip was re-magnetized, and the sample was negatively saturated. 
Phase-contrast MFM images were acquired at each OP field with tip-lift of 20 nm, and repeated $\sim$3 times over the same field-of-view to rule out tip-induced perturbations. 
\textbf{Quantitative MPC Analysis} was performed on a dataset acquired entirely using the same tip, with identical scan conditions across four samples for varying bias fields (\ref{fig:SAF-PhaseContrast}, SI $\S$S3). 
Great care was taken to avoid any changes to the MFM tip through the dataset, which would affect the MPC, and preclude quantitative comparisons. 
Similarly rigorous validation experiments were performed for multiple samples, including with co-located half- and full stacks (SI $\S$S4).
After processing the MFM images, spin-textures were individually identified and fit using established recipes to determine their MPC amplitude and width (\ref{fig:SAF-PhaseContrast}, SI $\S$S3). 

\textbf{Micromagnetic Simulations } were performed using the mumax$^3$ software package \cite{Vansteenkiste2014} with a 2 nm mesh over a 2 $\mathrm{\mu}$m-sized grid. 
The effective medium model \cite{Woo2016} was used to simulate reduced stacks with six magnetic layers L1-L6 (\ref{fig:SAF-MH_Sim}a), modelled upon experimental stacks. 
The simulated stacks used magnetic parameters from measurements (see SI $\S$S1) and related prior works on similar Ir/Fe/Co/Pt stacks \cite{Chen2022, Bottcher2023, Chen2023},with slight optimization of $K_{\rm eff}$. 
Established simulation recipes were used to generate the layer-wise magnetization profile with varying OP field \cite{Vansteenkiste2014, Tan2020, Chen2022}. \\
\textbf{IECs } between successive FMs were modelled using three distinct parameters to emulate different HM spacers – $\eta_{\rm 2}$ for Ir(1)/Pt(1); $\eta_{\rm 1}$ for Pt(0.8), and $\eta_{\rm FM/AF}$ for the Ir($t_{\rm Ir}$). 
Initial estimates for IECs used measured values for similar multilayers \cite{Legrand2020, Chen2023}. 
These values were optimized based on hysteresis loops, resulting in $\eta_1$ = 0.40 mJ/m$^2$, $\eta_2$  = 0.28 mJ/m$^2$, $\eta_{\rm FM}$= +0.0535 mJ/m$^2$ and $\eta_{\rm AF}$ = -0.535 mJ/m$^2$.